\documentclass[preprint,12pt]{elsarticle}
\usepackage{amssymb}
\usepackage{graphicx}

\journal{Physics Letters B}

\begin{document}

\title{
Evidence for $\alpha$-particle condensation in nuclei from the Hoyle state
deexcitation}

\author[ipno,ifin]{Ad. R. Raduta}
\author[ipno]{B.~Borderie}
\author[INFNSezioneCatania,FisicaCatania,Bologna]{E. Geraci}
\author[ipno,lpc]{N. Le Neindre}
\author[ipno]{P.~Napolitani}
\author[ipno]{M.~F.~Rivet}
\author[lns]{R.~Alba}
\author[lns]{F. Amorini}
\author[INFNSezioneCatania]{G. Cardella}
\author[saha]{M. Chatterjee}
\author[INFNSezioneCatania]{E. De Filippo}
\author[lyon]{D.~Guinet}
\author[lyon]{P.~Lautesse}
\author[INFNSezioneCatania,csf]{E. La Guidara}
\author[lns,kore]{G. Lanzalone}
\author[INFNSezioneCatania]{G. Lanzano\fnref{fn1}}
\author[lns,FisicaCatania]{I. Lombardo}
\author[lpc]{O.~Lopez}
\author[lns]{C.~Maiolino}
\author[INFNSezioneCatania]{A.~Pagano}
\author[INFNSezioneCatania]{S. Pirrone}
\author[INFNSezioneCatania,FisicaCatania]{G. Politi}
\author[lns,FisicaCatania]{F. Porto}
\author[lns,FisicaCatania]{F. Rizzo}
\author[lns,FisicaCatania]{P. Russotto}
\author[ganil]{J.P.~Wieleczko}
\fntext[fn1]{deceased}

\address[ipno]{Institut de Physique Nucl\'eaire, CNRS/IN2P3,
Universit\'e Paris-Sud 11, Orsay, France}
\address[ifin]{National Institute for Physics and Nuclear Engineering,
Bucharest-Magurele, Romania}
\address[INFNSezioneCatania]{INFN, Sezione di Catania, Italy}
\address[FisicaCatania]{Dipartimento di Fisica e Astronomia,
Universit\`a di Catania, Italy}
\address[Bologna]{INFN, Sezione di Bologna and Dipartimento di Fisica,
Universit\`a di Bologna, Italy}
\address[lpc]{LPC, CNRS/IN2P3, Ensicaen, Universit\'{e} de Caen, 
Caen, France}
\address[lns]{INFN, Laboratori Nazionali del Sud, Catania, Italy}
\address[saha]{Saha Institute of Nuclear Physics,
Kolkata, India}
\address[lyon]{Institut de Physique Nucl\'eaire, CNRS/IN2P3,
Universit\'e Claude Bernard Lyon 1, Villeurbanne, France}
\address[csf]{CSFNSM, Catania, Italy}
\address[kore]{Universit\`a di Enna ``Kore'', Enna, Italy}
\address[ganil]{GANIL, (DSM-CEA/CNRS/IN2P3), Caen, France}

\begin{abstract}
The fragmentation of quasi-projectiles from the nuclear reaction
$^{40}$Ca+$^{12}$C at 25 MeV/nucleon was used to produce excited states
candidates to $\alpha$-particle condensation. 
Complete kinematic characterization of individual decay events,
made possible by a high-granularity 4$\pi$ charged particle multi-detector,
reveals that 7.5$\pm$4.0 \% of the particle decays of the Hoyle state
correspond
to direct decays in three equal-energy $\alpha$-particles. 
\end{abstract}



\maketitle


Bose-Einstein condensation is known to occur in weakly and strongly
interacting systems such as dilute atomic gases and liquid
$^4$He~\cite{BEC}.
During the last decade it was theoretically shown that
dilute symmetric nuclear matter may also experience Bose particle
condensation \cite{ropke_nm_prl1998,beyer_plb2000,sogo_prc2009}.
More precisely, for densities smaller than one fifth of the 
nuclear saturation density, nuclear matter organizes itself in 
$\alpha$-clusters, while for higher densities the 2-nucleon 
deuteron condensation is preferred.
This new possible phase of nuclear matter may have its counterpart in
low-density states of self conjugate lighter nuclei, in the same way
as superfluid nuclei are the finite-size counterpart of 
superfluid nuclear and neutron matter.
This means that under some circumstances,
the alpha condensation, i.e. bosonic properties, 
might dominate over the nucleon properties even in finite nuclei.


Considerable advance in favour of  $\alpha$-particle condensation
in nuclei is provided by the excellent theoretical description
of the Hoyle state ({\em i.e.} the first 0$^+$ excited state  at
7.654 MeV of $^{12}$C) and of the 0$_6^+$ state at 15.097 MeV of $^{16}$O 
in terms of condensate type wave functions
\cite{tohsaki_prl2001,funaki_prl2008}.
The case of the Hoyle state is particularly suggestive 
as both shell-model and no-core shell model calculations 
are known to fail in describing it.
Recent Fermionic Molecular Dynamics (FMD) calculations \cite{fmd}
plead in favour of a more nuanced interpretation of the Hoyle state structure,
that is a mixture of various pre-formed $\alpha$-configurations.
This scenario agrees with the pioneering works of Uegaki {\it et al.} 
\cite{uegaki77} where acute-angle, bent and linear chain configurations 
were identified. 
In what regards the expected diluteness, 
coupled channel analysis \cite{ohkubo_prc2004} 
of experimental data confirm
theoretical works \cite{tohsaki_prl2001} 
estimating the rms radius of the Hoyle state as 45\% larger 
than the radius of $^{12}$C in its ground state.
More generally,  $\alpha$-particle condensation is conjectured
to be a generic feature of medium-size self-conjugated 4$N$ nuclei 
whose excitation lies in the vicinity of the $N\alpha$ decay threshold
\cite{schuck2004,WvO2006}.


The aim of the present Letter is to search from the experimental
side a direct evidence for $\alpha$-particle condensation from
the Hoyle state.
According to the present understanding of the Hoyle state: a gas-like
structure of three $\alpha$-particles which occupy dominantly the lowest
S orbit, such an evidence
may be judged from the simultaneous emission of three
$\alpha$-particles with very
low kinetic energy dispersion.
As it is experimentally difficult to identify and measure
$\alpha$-particles with
low energy, probably the most appropriate strategy, chosen in this work,
should involve high velocity
reaction products in the laboratory (to take advantage of velocity boosts)
detected by a
high granularity-high solid angle particle array (to precisely reconstruct
the directions of velocity vectors).
The existence of the direct decay channel was firstly investigated in
Ref. \cite{Fre94} and its contribution to the total alpha decay
(dominated by the $^8$Be$_{g.s.}$ + $\alpha$ sequential decay) was 
estimated to be at most 4\%. 
The existence of such a decay channel is also of the
greatest importance for nuclear astrophysics, due to the crucial role
played by the Hoyle state for the synthesis of $^{12}$C in the
universe~\cite{Hoyle53,Cook57}.


The data reported here have been obtained in the nuclear reaction 
$^{40}$Ca+$^{12}$C at 25 MeV per nucleon incident energy performed at INFN,
Laboratori Nazionali del Sud in Catania, Italy. 
The beam impinging on a thin carbon target (320 $\mu$g/cm$^2$)
was delivered by the Superconducting Cyclotron and the
charged reaction products were detected by the CHIMERA 4$\pi$
multi-detector \cite{chimera}. The beam intensity was kept around
$10^7$ ions/s to avoid pile-up events.
CHIMERA consists of 1192 silicon-CsI(Tl) telescopes mounted on 35
rings covering 94\% of the solid angle, with polar angle ranging from
1$^{\circ}$ to 176$^{\circ}$. 
The solid angle corresponding to each module is not uniform, 
but varies between 0.13 msr
at forward angles and 35.4 msr at the most backward angles. 
Among the most interesting characteristics of CHIMERA are 
the low detection and identification thresholds
for light charged particles (LCP)
and the very high granularity at forward angles.  
The mass and charge of the detected
nuclei were determined by the energy-time of flight method
(TOF) for LCP stopped in silicon detectors and
$\Delta E-E$ ($Z>5$) and
shape identification ($Z \leq 5$) techniques
for charged products stopped in CsI(Tl).
In addition $^8$Be nuclei (two equal-energy $\alpha$'s 
hitting the same crystal) were identified in CsI(Tl) \cite{Mor10}.
The energy of detected nuclei was measured by the Si detectors
calibrated using proton, carbon and oxygen beams
at various energies ranging from 10 to 100 MeV. 
For $Z=2$, dedicated energy calibrations of the fast component of CsI(Tl)
light was realized using the TOF.
Though high-quality TOF charts allowing for a direct calibration of the
CsI light
exist only for 60\% of the total number of modules,
we have finally managed to calibrate more than 95\% of modules
from 1$^{\circ}$ to 62$^{\circ}$.
The modules for which TOF information was poor or missing were calibrated
by comparing the fast component distribution 
with the benchmark distribution of the corresponding ring built out 
of the telescopes with excellent TOF.
The energy resolution for alpha particles varies between 1.0 and 2.5\%
depending on the module.

\begin{figure}
\begin{center}
\includegraphics[angle=0, width=0.6\columnwidth]{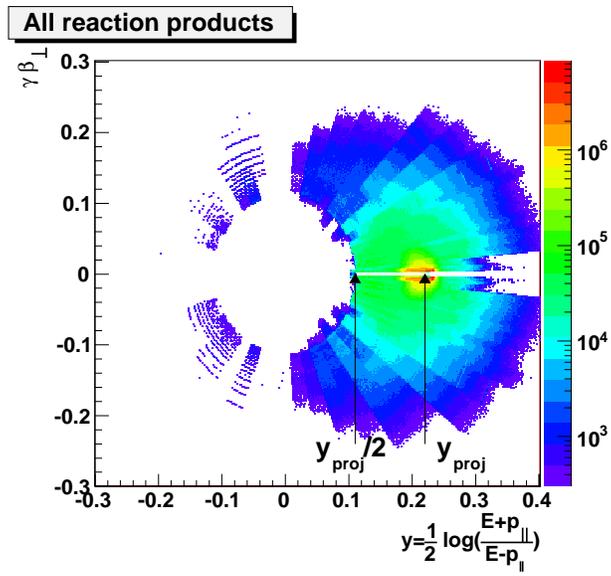}
\end{center}
\caption{(Color online)
Invariant velocity plot corresponding to all reaction products 
obtained in $^{40}$Ca+$^{12}$C at 25 MeV/nucleon.
The considered events have $m_{\alpha} \geq 3$.
The vertical arrows indicate the projectile rapidity and its half.}
\label{fig:vparvper}
\end{figure}

Complete kinematical characterization of the 
reaction products allows one to identify the particle emitting 
sources using invariant velocity plots in the 
$\gamma \beta_{\perp}$ vs. $1/2 \log\left[\left(E+p_{\parallel}\right)/
\left(E-p_{\parallel}\right)\right]$ plane.
$E$, $p_{\parallel}$, $p_{\perp}$ stand for the energy and momentum components
along and, respectively, perpendicular to the beam axis,
$\beta=v/c$ is the reduced velocity and
$\gamma=1/\sqrt{1-\beta^2}$.
The invariant velocity plot corresponding to events having
an $\alpha$-multiplicity of at least 3
is depicted in Fig. \ref{fig:vparvper}.   
It shows the dominant binary character of the collisions 
with the formation of a quasi-projectile (QP) and a quasi-target (QT)
not studied in this experiment.  
The QP source may be easily isolated by requiring a 
rapidity of reaction products larger than $y_{proj}/2$,
where  $y_{proj}$ is the projectile rapidity.
Hereafter, we shall focus exclusively on QP decay products with $m_{\alpha} = 3$.

\begin{figure}
\begin{center}
\includegraphics[angle=0, width=0.45\columnwidth]{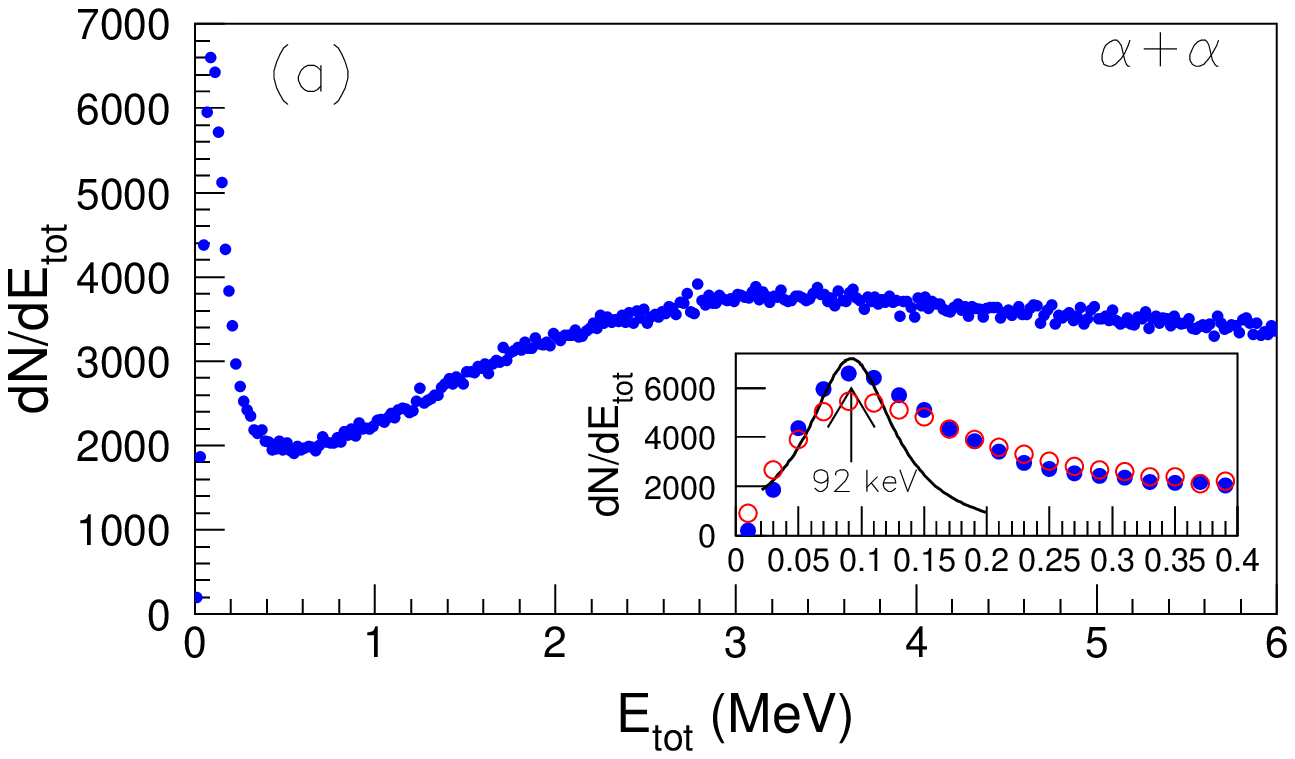}
\includegraphics[angle=0, width=0.45\columnwidth]{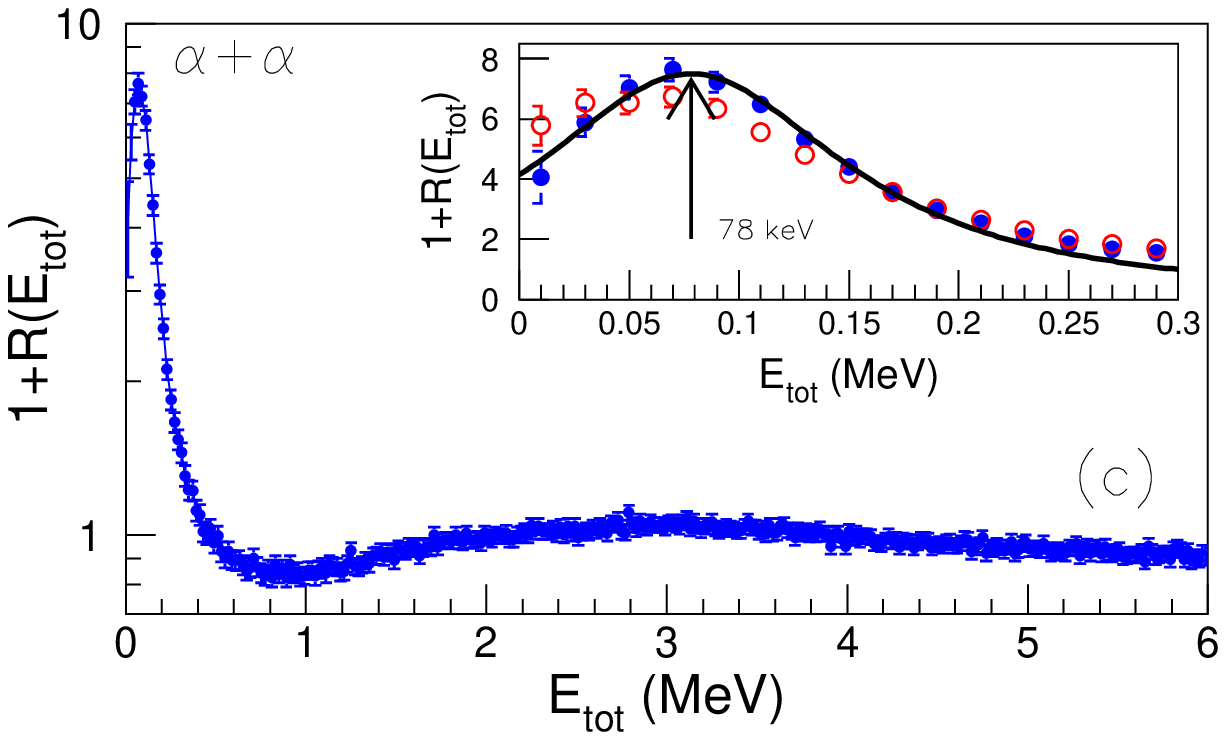}
\includegraphics[angle=0, width=0.45\columnwidth]{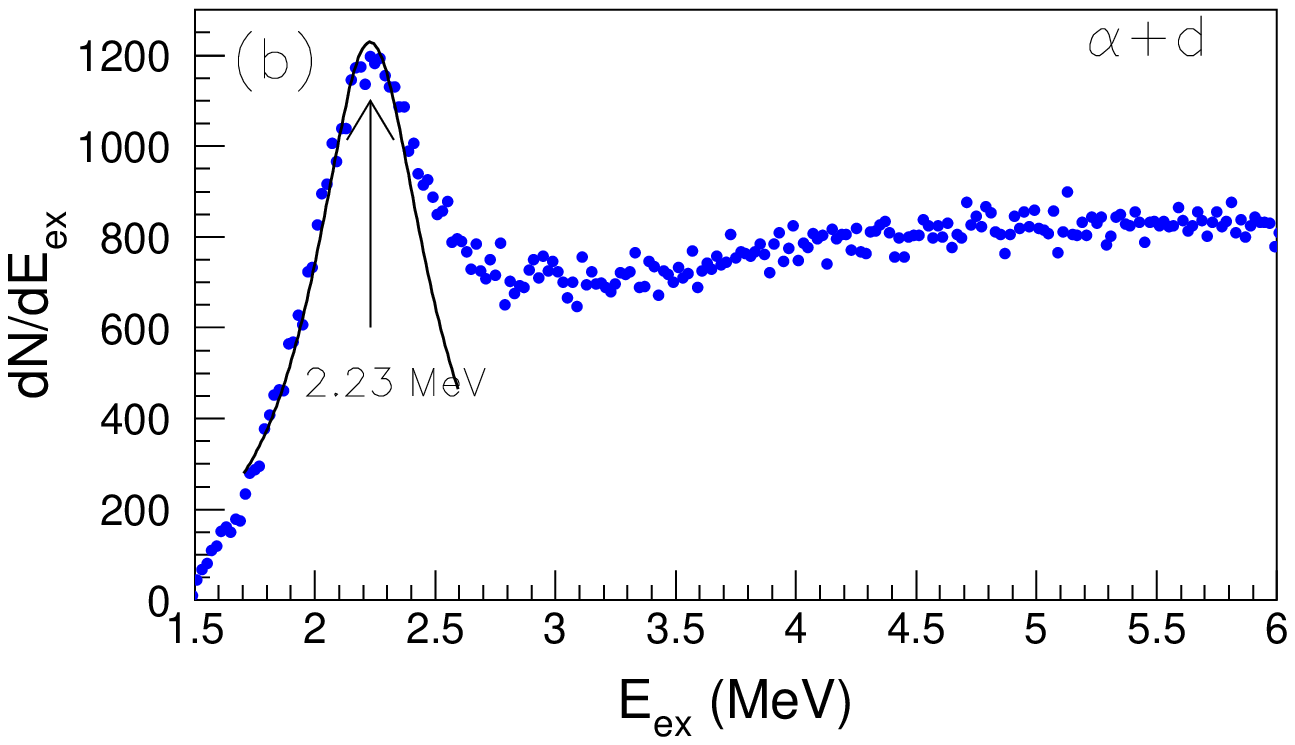}
\includegraphics[angle=0, width=0.45\columnwidth]{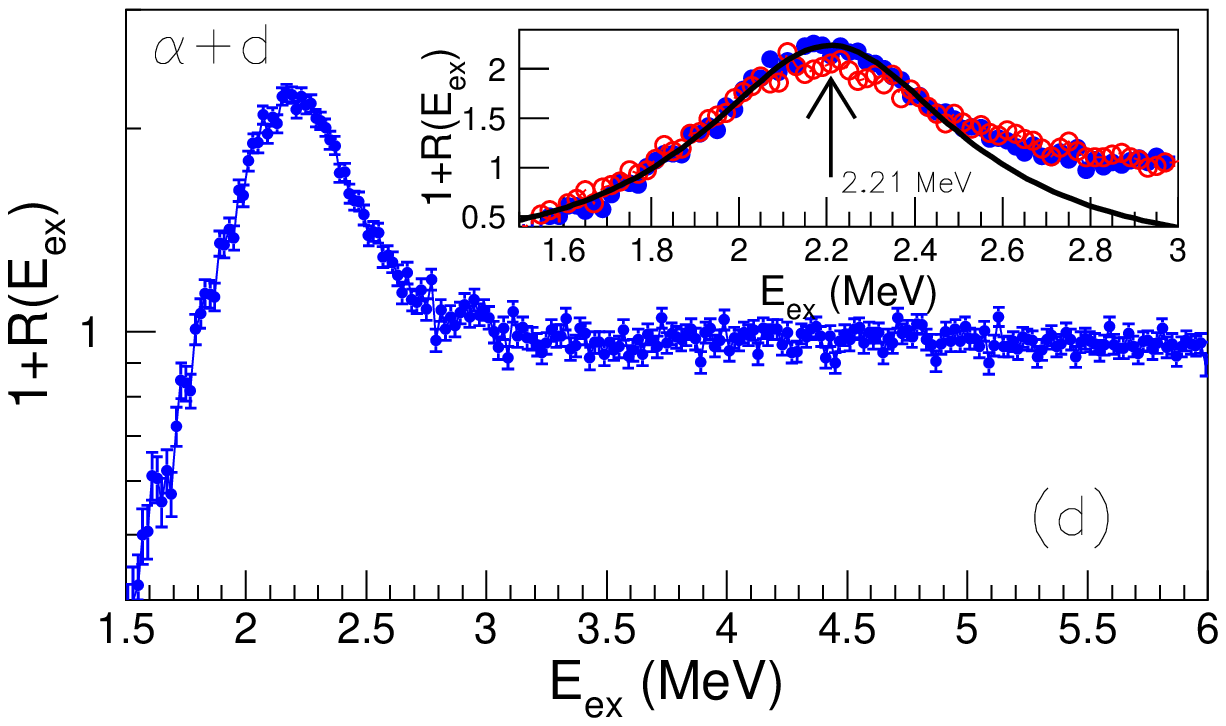}
\end{center}
\caption{(Color online)
Yields of correlated $\alpha-\alpha$ (a) and
 $\alpha$-d (b) emissions out of QP expressed 
as a function of total kinetic energy (a) or 
excitation energy (b) and corresponding correlation functions (c and d). 
The insets in (a) and (c) correspond to zooms on the $^8$Be(g.s.) peak.
The inset in (d) details the $\alpha$-d correlation function 
in the domain of $^7$Li(2.19 MeV).
Peak fits using Breit-Wigner distributions are illustrated 
with solid lines, while centroids are pointed by arrows.
Full and open symbols in the insets correspond to different
ways of calculating the angle
under which the detected particles were emitted (see text). 
}
\label{fig:test_calib}
\end{figure}

With the aim of investigating the reliability of the energy calibration,
we show in
Fig. \ref{fig:test_calib} the $\alpha-\alpha$ (top panel-left) and
$\alpha$-d (bottom panel-left) correlated spectra as a function of 
total kinetic energy in the center-of-mass of the two particles (CM), 
$E_{tot}=\sum_i E_i^{(CM)}$, and, 
respectively, excitation energy ($E_{ex}=E_{tot}-Q$).
For $\alpha-\alpha$ correlation we considered the QP event sample 
described above.
For $\alpha$-d, we have additionally set the deuteron multiplicity
to at least 1.
The solid lines correspond to the peak fits using a Breit-Wigner
distribution and peak centroids are pointed out by arrows.
One may see that the $\alpha-\alpha$ spectrum shows a 
narrow peak centered at 92 keV ($\Gamma$=84 keV) and 
a much broader peak centered around 3 MeV.
The first peak corresponds to the ground state of $^8$Be ($Q$=-92 keV) 
with $\Gamma^{exp}$=5.57 eV and the second one to the first excited state 
at 3.03 MeV ($\Gamma^{exp}$=1.5 MeV). 
The $\alpha$-d spectrum shows a well-formed peak centered at 2.23 MeV
($\Gamma$=570 keV) corresponding to the first excited state of
$^6$Li at 2.186 MeV ($\Gamma^{exp}$=24 keV).
In both cases, the low energy levels are accurately determined
(within 40-50 keV), 
meaning that the charged particle energy calibration is trustworthy.
At variance with this, information on higher excited levels 
is blurred by particles originating from other decays whose
abundance increases with energy.
Nevertheless, contamination of spectra may be removed to a large extent 
by employing correlation function techniques 
which account for how much the correlation within the physical event 
differs from the underlying single-particle phase space. 

Correlation functions (CF) are defined as the ratio between 
the correlated (physical) yield  $Y_{corr}$ 
and the product of single particle yields, generically termed
as uncorrelated spectrum $Y_{uncorr}$, measured under the same conditions,
\begin{equation}
1+R(X)=\frac{Y_{corr}(X)}{Y_{uncorr}(X)}.
\label{eq:corrf}
\end{equation}
$Y_{uncorr}$ can also be built by mixing particles from different events
as it is done in this work.
In multi-particle CF~\cite{charity_prc1995},
the generic variable $X$ is represented by
the total kinetic energy of the particles of interest in their center-of-mass
frame $E_{tot}$, by the excitation energy of their emitting source/state, 
$E_{ex}=E_{tot}-Q$ or, in the case of two-particle correlations, by the
relative momentum.
In nuclear physics, CF have been exploited 
to access spectroscopic properties \cite{tan},
give space-time information taking advantage of
proximity effects induced by Coulomb repulsion, and
emphasize any production of events or sub-events with
specific rare partitions
\cite{charity_prc1995,Kim92,gabi_epja2003}.
The flat shape of the $\alpha$-d CF 
around 1+R(X)=1 at $E_{ex}>$3 MeV (see (d) of Fig. \ref{fig:test_calib}) 
confirms that the uncorrelated spectra are under control. 

Both correlated spectra and CF manifest
a peak broadening. This is a genuine 
consequence of detector finite granularity and energy resolution.
Numerical simulations confirm that experimental distributions are
compatible with an average energy resolution of detected alpha
particles $R_E$ of 2\%.
The finite granularity is also responsible for a certain imprecision in the
determination of the energy.
To illustrate this, the insets in Fig. \ref{fig:test_calib}
(a), (c) and (d)
confront the correlated spectra and CF obtained under different hypotheses 
on the angle under which the detected particles were emitted. 
In the first case (solid symbols) we considered that all the particles 
which hit a certain module have been emitted under the
angle corresponding to the geometrical center of that module. 
An alternative solution is to attribute to each
particle a random angle in the domain allowed
by the geometrical extension of the detector (open symbols). 
For all considered cases, the two scenarios lead to similar results,
though the first hypothesis is preferable as
producing smoother distributions and no extra broadening. 
In what follows only results obtained under the first hypothesis
will be presented.

\begin{figure}
\begin{center}
\includegraphics[angle=0, width=0.7\columnwidth]{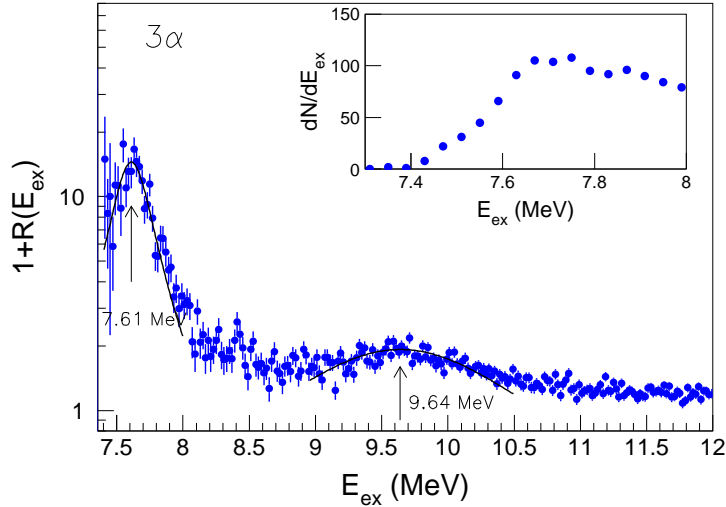}
\end{center}
\caption{(Color online)
 3-$\alpha$ correlation function as a function of excitation energy.
The arrows correspond to centroids of Breit-Wigner distributions 
(solid lines).
Inset: zoom of the correlated spectrum in the energy domain 
of the Hoyle state.
}
\label{fig:corrf}
\end{figure}

Information on the $\alpha$-particle unstable excited states of $^{12}$C
populated by the $^{40}$Ca+$^{12}$C at 25 MeV/nucleon reaction
may be extracted from the 3$\alpha$-CF. 
The CF plotted in Fig. \ref{fig:corrf}
corresponds to the whole set of events with $m_{\alpha}=3$
and an uncorrelated spectrum built by full event mixing 
({\it i. e.} each particle belongs to a different physical event). 
The error bars are calculated
considering only the statistical errors of the correlated spectrum.
Errors on the uncorrelated spectra have been reduced to negligible
values by increasing the number of uncorrelated events as
compared to correlated events. 
The CF shows two peaks centered at $E_{ex}$=7.61 MeV ($\Gamma$=0.33 MeV) 
and $E_{ex}$=9.64 MeV ($\Gamma$=1.14 MeV). 
The first peak corresponds to
the Hoyle state ($E_{ex}^{exp}$=7.654 MeV, $\Gamma^{exp}$=8.5 eV), 
while the second one is due to the complex region of excitations,
characterized by the strong $E_{ex}^{exp}$=9.64 MeV ($\Gamma^{exp}$=34 keV), 
$3^-$ state and
by the broad $E_{ex}^{exp}$=10.3 MeV, $0^+$ 
state submerging a possible $2^+$ state at 9.7 MeV
\cite{itoh_npa2004,freer_npa2010}.
As before, the peak centroids reproduce well the experimental excitation 
energy values, while a significant broadening comes from 
the non-ideal detector properties 
(granularity and energy resolution).
The inset of Fig. \ref{fig:corrf} illustrates
the correlated spectrum in the energy domain of the Hoyle state and
allows to estimate the relative amount of genuine 3$\alpha$ decays 
versus background events. 
In the domain 7.37-7.97 MeV this ratio amounts to 1.

\begin{figure}
\begin{center}
\includegraphics[angle=0, width=0.99\columnwidth]{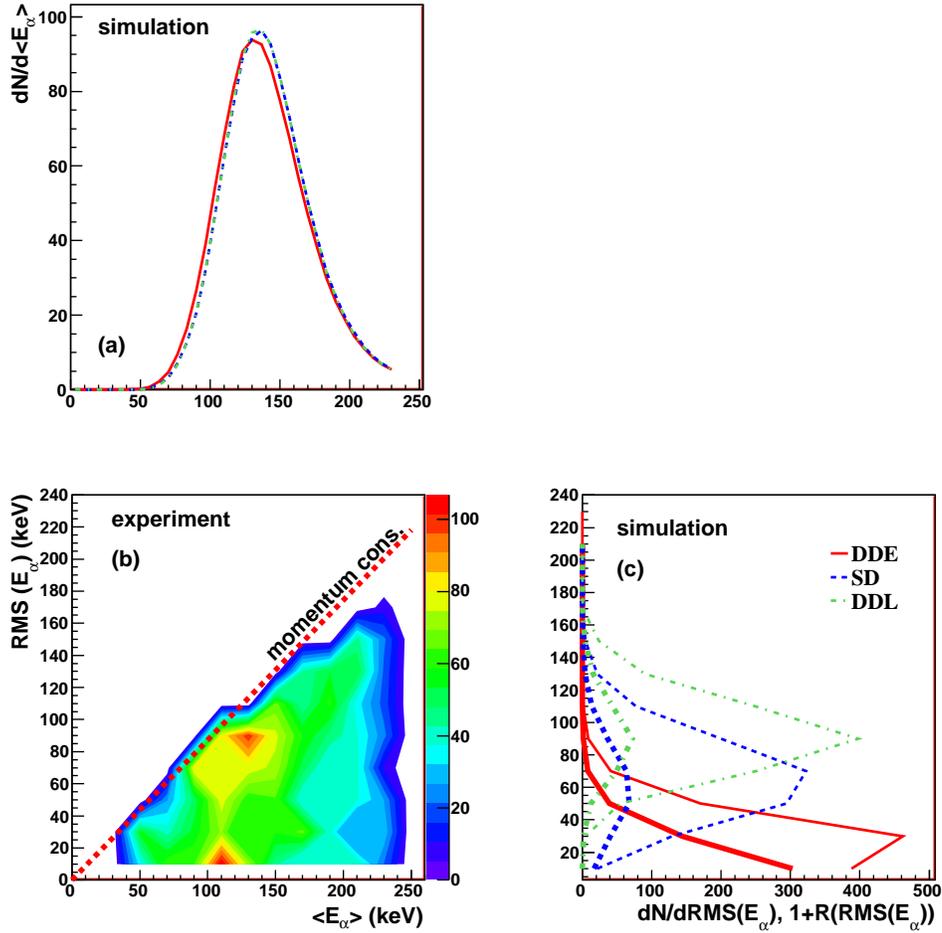}
\end{center}
\caption{(Color online)
Experimental three-$\alpha$ correlation function (b)
expressed as a function of average kinetic energy - 
RMS of $\alpha$ particles 
corresponding to 1072 experimental events with
$7.37 \leq E_{ex} \leq 7.97$ MeV.
The uncorrelated yield is built such as to allow for decay through $^8$Be.
The dotted line marks the maximum RMS compatible with momentum
conservation.
$\langle E_{\alpha}\rangle$ (a) and RMS (c) spectra (normalized to 1072 
events) of simulated 
DDE (solid lines), SD (dashed lines) and DDL (dot-dashed lines) decays 
of the Hoyle state after filtering 
through the detector replica with $R_E$=2\%.
Panel (c) presents also the RMS projection of 
$Y_{corr}(\langle E_{\alpha}\rangle,RMS)/Y_{uncorr}(\langle E_{\alpha}\rangle,RMS)$ 
(thick lines).
}
\label{fig:cfiv}
\end{figure}

We demonstrated so far that the $^{40}$Ca+$^{12}$C 
nuclear reaction at 25 MeV/nucleon
populates excited states of $^{12}$C
nuclei which decay by 3-$\alpha$ emission.
We now search for possible direct decays of the Hoyle state into
equal energy particles (DDE) and,
eventually, estimate their branching ratio.
The existence of DDE is manifest when looking at a two-dimensional
CF in the $\langle E_{\alpha}\rangle$-RMS plane
(Fig. \ref{fig:cfiv}(b));  
$\langle E_{\alpha}\rangle$ is the average
kinetic energy of $\alpha$-particles in their CM reference frame and
RMS=$\sqrt{\langle E_{\alpha}^2\rangle-\langle E_{\alpha}\rangle^2}$.
The uncorrelated spectrum is built by partial event mixing
~\cite{grenier_npa2008} of experimental events
by taking two particles from 
the same event, in order to mimic for decay through $^8$Be,
while the third one stems from a different event.
The peak of the correlation function localized at
$\langle E_{\alpha}\rangle$ = 110 keV and very low
RMS$\leqq$ 25 keV corresponds
to an equal sharing of the available energy of the Hoyle state
among the three $\alpha$ particles. Note that 110 keV corresponds
to the value of the maximum of the CF of Fig.~\ref{fig:corrf}:
110 keV $\simeq$(7610+Q)/3.
To go further in the interpretation of the CF and estimate 
branching ratios, the experimental results were compared to
numerical Monte-Carlo simulations 
filtered by the multi-detector replica.
These are performed by considering  
$^{12}$C nuclei excited in the Hoyle state and boosted with a velocity
distribution identical to the experimental one. They decay by one
of the following mechanisms: 
(1) direct emission in three $\alpha$-particles with equal energies (DDE),
(2) sequential decay proceeding via the g.s. of $^8$Be with isotropic
emission of the two $\alpha$-particles from $^8$Be (SD) and
(3) direct $\alpha$ decay from a linear chain (DDL)~\cite{matsu2004,epel2011} 
with an
$\alpha$ at rest and an equal sharing of the available energy of the Hoyle
state between the two other $\alpha$ particles.
Limitation of direct decays (DD), 
which in principle assume random energy sharing
among the emitted particles, to the particular sub-class of DDE is
due to the predicted condensate nature of the Hoyle state.
In each case and, for sequential decay, at each decay step
the available energy ($E_{ex}+Q$) is shared among the reaction products
so as to conserve the linear and angular momenta and, eventually,
obey the postulated particularities of the decay.
No preferential orientation between emitted particle velocities and 
source boost exists.
The obtained reaction products are assumed to freely
propagate toward the detector. 
Once the detector is reached, the events are filtered: 
if one or more particles hit a dead detector 
area or a detector which is out of work, the event is suppressed.
If all particles have been detected, 
the velocity vectors are altered such as
to account for finite angular and energy resolution.
The simulated events are then analyzed as the experimental ones.
The obtained detection efficiencies 
 vary from 41 to 49\% depending on the decay type.
A comparison of filtered and normalized DDE (full), SD (dashed) and 
DDL (dot-dashed) simulations is depicted with thin lines in
Fig. \ref{fig:cfiv} as a function of $\langle E_{\alpha}\rangle$ (a) 
and RMS (c).
Very little sensitivity on the decay mechanism
of $Y_{corr}(\langle E_{\alpha}\rangle)$
is observed, which is due to the angular resolution of the 
detection modules.
By contrast, the kinetic energy dispersion (RMS) in the emitter CM
manifests measurable sensitivity to the decay mechanism.
For the CHIMERA granularity and a perfect energy resolution
$Y_{corr}(RMS)$ are peaked at 10, 70 and 90 keV for DDE, SD and, 
respectively, DDL while for an average resolution $R_E$=2\%  
the corresponding values are 30, 70 and 90 keV.
\begin{figure}
\begin{center}
\includegraphics[angle=0, width=0.60\columnwidth]{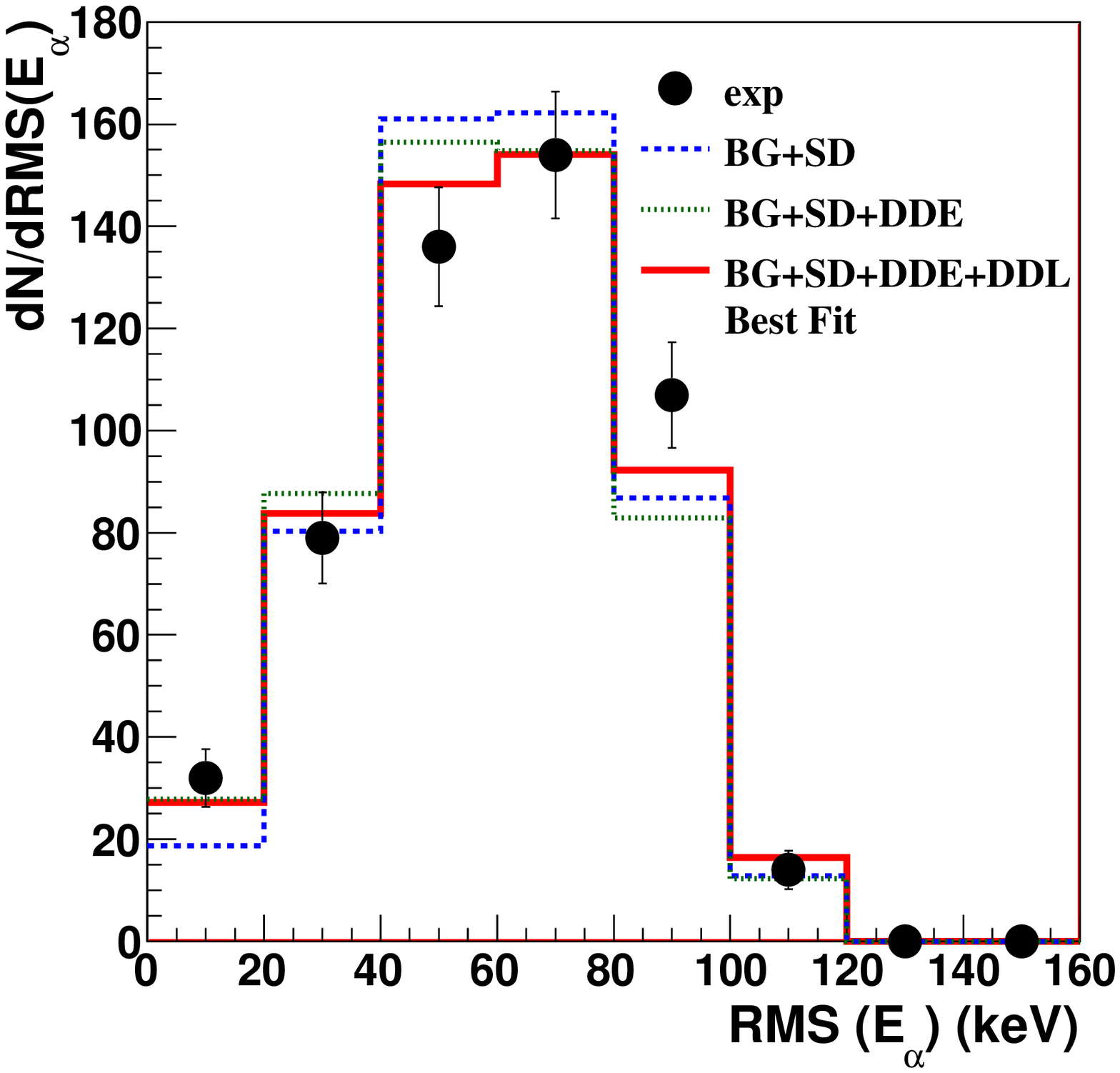}
\end{center}
\caption{(Color online)
$RMS$ spectra of experimental (solid points) and best
$\chi^2$ simulated events (solid line) 
in the region $90 \leq \langle E_{\alpha}\rangle \leq 150$ keV. 
Experimental error bars are statistical.
Dashed and dotted histograms illustrate simulation results if only
sequential decay (60\%) and, respectively, sequential decay (55.5\%) + 
DDE (4.5\%) are considered.
In all cases, background (BG) events amount to 40\%.  
}
\label{fig:bestfit}
\end{figure}
This suggests that searching for the best agreement between
experimental and
simulated $Y_{corr}(\langle E_{\alpha}\rangle,RMS)$ constitutes 
a pertinent procedure to quantify each decay channel.
Not surprisingly, the general category of DD with 
random energy sharing (not considered hereafter) leads
to a broad $Y_{corr}(RMS)$ distribution which, 
under our experimental conditions, looks very similar to the 
dominating SD~\cite{Fre94}.
For a realistic reproduction of the experimental data, we add to the
pure simulated events the same proportion of background events that 
exists in the data. 
These are produced by partial event mixing~\cite{grenier_npa2008}
of experimental events.
The comparison is restricted to the energy domain,
$7.4 \leq E_{ex} \leq 7.8$ MeV, to reduce the importance of background events
to 40\%.
Applying a $\chi^2$ minimization procedure and correcting for efficiencies, 
we infer that 7.5$\pm$4.0\% of events correspond to DDE, 
9.5$\pm$4.0\% to DDL and 83.0$\pm$5.0\% to SD. 
Error bars are estimated by taking into account
statistical, $\chi^2$ and background errors.
The RMS($E_{\alpha}$) spectrum of simulated events (solid line)
corresponding to the best $\chi^2$ and the energy domain
$90 \leq \langle E_{\alpha}\rangle \leq 150$ keV
is displayed in Fig. \ref{fig:bestfit} 
and compared to the corresponding experimental data (solid points). 
Note that a limitation to DDE and SD as possible decays (dotted histogram)
leads to a significant increase of $\chi^2_{min}$ without changing
the DDE percentage. 
Results of simulations in which one allows only for SD are plotted
with dashed lines for the sake of completeness. As one may notice, they
correspond to a worse agreement with the data, particularly
a sizeable underestimation
of the number of low-RMS events. 
In all cases, the amount of background events is 40\%.

Finally we can now fully understand the CF
pattern (Fig. \ref{fig:cfiv}(b)); 
panel (c) of the figure presents also the simulated behaviours of 
$Y_{corr}(RMS)/Y_{uncorr}(RMS)$ (thick lines).
Apart from DDE,
the broad region around $\langle E_{\alpha}\rangle$ = 90-130 keV and 
centered at RMS$\approx$ 70 keV corresponds to
the sharing of the available energy between the two 
$\alpha$'s of $^8$Be and the remaining $\alpha$ of 191 keV.
The peak at $\langle E_{\alpha}\rangle$$\approx$ 130 keV and
RMS = 90keV corresponds to the direct decay of a linear chain.
Though the CF shows, as expected, a certain sensitivity to the event 
selection or event mixing recipe, its overall pattern remains stable.
Particularly the DDL peak of the CF
is systematically shifted by about 20 keV with respect to the DDE one.
As early mentioned, a small shift of the correlated spectrum is
attributed to the finite
size of detection modules (see Fig. \ref{fig:cfiv}(a)). An extra shift
could be introduced by statistical effects.

A more popular way to visualize competing 3-particle decay mechanisms 
is the Dalitz plot. 
In this representation, DDE events must concentrate around the origin.
Figure 1 in reference~\cite{adriana2011} shows that our data manifest this
pattern. Note also that, because of limited statistics and background,
the rest of the figure cannot be interpreted.

The same analyses have been performed for the
complex region centered at 9.64 MeV where the statistics is much higher; 
no indication in favour of a direct 3-$\alpha$ decay with equal
energies has been
obtained~\cite{adriana2011}.
For the 0$_6^+$ state at 15.097 MeV of $^{16}$O
the statistics is too poor to allow unambiguous interpretation of
low RMS events.

In conclusion, the nuclear reaction $^{40}$Ca+$^{12}$C at 
25 MeV/nucleon bombarding energy was used to produce states 
theoretically predicted as $\alpha$-particle condensate states.
Supposing that equal values of kinetic energy of the emitted
$\alpha$-particles represent a sufficient criterion for establishing
the existence of $\alpha$-particle condensation,
we found that 7.5$\pm$4.0\% of events corresponding to the 
Hoyle state decay fulfil this criterion. 
To our knowledge, this is the first direct experimental indication of 
$\alpha$-particle condensation in nuclei.
This study also evidenced
the presence of direct alpha decays from a linear
$\alpha$-chain as recently theoretically mentioned~\cite{matsu2004,epel2011}.
An experiment with higher statistics is planned to study
the $^{16}$O case.

Acknowledgments.
The authors are indebted to P.~Schuck for numerous discussions 
and one of the authors Ad. R. R. acknowledges the partial financial support 
from ANCS, Romania, under grant Idei nr. 267/2007.


\end{document}